\documentclass[12pt]{iopart}
\usepackage{latexsym,amssymb,amsbsy,graphicx}
\usepackage[space]{cite} % If exist.
\usepackage{xcolor}
\newcommand{\noi}{\noindent}
\newcommand{\dint}{\displaystyle\int}
\newcommand{\lam}{\lambda}
\newcommand{\fr}{\frac}
\newcommand{\hb}{\hbar}

\newcommand{\vk}{{\bf k}}
\newcommand{\vr}{{\bf r}}

\newcommand{\akl}{b_{\vk,\lam}}
\newcommand{\akla}{b^+_{\vk,\lam}}

\newcommand{\om}{\omega}
\newcommand{\Om}{\Omega}

\newcommand{\bg}{\begin{equation}}
\newcommand{\en}{\end{equation}}
\newcommand{\dsum}{\displaystyle\sum}
\newcommand{\nn}{\nonumber}

\newcommand{\lan}{\langle}
\newcommand{\ran}{\rangle}
\newcommand{\Gam}{\Gamma}
\newcommand{\ba}{\begin{eqnarray}}
\newcommand{\ea}{\end{eqnarray}}
\newcommand{\vekl}{\hat{\bf\varepsilon}_{\vk}^{(\lam)}}
\newcommand{\veps}{{\bf\varepsilon}}

\begin{document}

%%%%%%%%%%%%%%%%%%%%% Publisher's Area please ignore %%%%%%%%%%%%%%
%\catchline{}{}{}{}{}
%%%%%%%%%%%%%%%%%%%%%%%%%%%%%%%%%%%%%%%%%%%%%%%%%%%%%%%%%%%%%%%%%%%

\title[Fluorescence by a polar quantum system in a polychromatic field]{Fluorescence by a polar quantum system in a polychromatic field}

\author{N.N.\,Bogolyubov, Jr.}

\address{Dept. of Mechanics, V.A. Steklov Mathematical Institute
 of the RAS, 8, Gubkina str.,
Moscow, 119991, Russia\\ bogolubv@mi-ras.ru}

\author{A.V.\,Soldatov}
\address{Dept. of Mechanics, V.A. Steklov Mathematical Institute
 of the RAS, 8, Gubkina str.,
Moscow, 119991, Russia\\
soldatov@mi-ras.ru}

\vspace{0.3cm}

\begin{abstract}

Spectral properties of fluorescent radiation from a two-level
quantum system with broken inversion spatial symmetry, which can
be described by a model of an one-electron two-level atom whose
electric dipole moment operator has permanent unequal diagonal
matrix elements, were studied.   The case of the excitation of
this system by a polychromatic laser field, comprised of $N-1$
high-frequency  components  with the frequencies close to or being
in resonance with the atomic transition frequency, and a
low-frequency component whose frequency coincides with the Rabi
frequency of the high-frequency components, was considered.
Special attention was given to the resonant bichromatic and nearly
resonant trichromatic excitation. In the former case it was shown
that by changing the intensity of the low-frequency component, one
can efficiently alter spectral properties of the fluorescent
radiation of the system in the high-frequency range, while in the
latter case it was found that the fluorescent behavior of the
system in question reveals a kind of an optoelectronic transistor
effect. Options for the experimental detection and practical usage
of the effects under study were discussed.

\end{abstract}


\markboth{Fluorescence by a Polar Quantum System ...} { }


 ======================================
\maketitle

\section{Introduction}

Resonance fluorescence, i.e. the phenomenon caused by resonant or
nearly resonant interaction of an atomic system  with
electromagnetic radiation, has been given a lot of attention since
its original theoretical prediction by B.R. Mollow \cite{Mollow1},
who showed that the typical Lorentzian-shape spectrum of the
spontaneous radiation emitted by an excited two-level atom into
the vacuum can be altered to a great degree by driving the atom
with a strong enough external monochromatic field with its
frequency $\om_1$ being in approximate resonance
 with the atomic radiation transition frequency $\om_0$. The characteristic
  Mollow triplet spectrum arises when the atomic spontaneous emission rate $\Gamma$
   is much smaller than the Rabi frequency $\Om^{(1)}_R$ of the driving field,
    which is in its turn smaller than the energy
spacing of the two-level atom, i.e. $\Gamma <<\Om^{(1)}_R<<
\om_0$. This altered spectrum is made of the prominent Lorentzian
spectral peak with the  natural linewidth $\Gamma$ at the resonant
atomic frequency accompanied by two sidebands being $1/3$ as high
as the central peak and shifted from that peak by the amount equal
to the Rabi frequency $\Om^{(1)}_R$ of the driving field. The
sidebands were of the Lorentzian shape too, but 1.5 times wider
than the central peak (see Fig.\ref{f1tor}).
 It was also found that the ratios of the widths and heights of the
  central peak to the ones of the sidebands remain virtually
  unaltered within a broad enough
range of the driving field strength. These theoretical predictions
by Mollow were confirmed experimentally
later\cite{Schuda,Wu,Hartig}. A sketch of a typical setup for such
experiments can be found, for example, in \cite{Scully}. By now,
resonance fluorescence developed into a useful analytical
technique in modern spectroscopy because the study of spectral
properties of the radiation field scattered by various quantum
system is fundamental to a number of research disciplines in
optics and laser physics and provides a convenient opportunity for
rigorous examination of the intrinsic characteristics of these
systems and the mechanism of their interaction with the radiation
field. Also, this phenomenon can be employed as a powerful tool
for the control of various quantum systems, naturally occurring
atoms, molecules as well as artificially designed quantum
nanosystems being among them. However, this important quantum
optical effect has not been sufficiently analyzed in polar quantum
systems, in which broken spatial inversion symmetry induces new
effects due to the presence of the permanent electric dipole
moments supported by polar systems. Unique optical properties of
polar quantum systems has already stimulated research efforts in
order to employ them as various radiation sources, see, e.g.
\cite{Izadnenas2023} and refs. [1-12] therein. In particular, the
interaction of the permanent dipole moment with external exciting
electromagnetic fields adds new features to the resonance
fluorescence phenomenon. Among them, the ability of a polar
two-level quantum system to fluoresce at much lower frequency when
being driven by high-frequency resonant monochromatic field was
predicted in \cite{Kibis:2009}. Spectral properties of its
low-frequency radiation and their dependence on the parameters of
the system and exciting field  were thoroughly studied later for
the case of a monochromatically driven two-level polar system
damped, in addition, by a vacuum dissipative
reservoir\cite{Soldatov:2016,Bogolyubov:2018}. It was also shown
that the same system can amplify the weak probe low-frequency
radiation\cite{Soldatov:2017}.  The case of the resonance
fluorescence in a polar two-level system interacting with a
broadband squeezed vacuum dissipative reservoir was studied for a
weak driving monochromatic field\cite{BSPEPAN:2020,BSJP:2020}, and
the case of interaction with a finite band degenerate and
non-degenerate  squeezed vacuum dissipative reservoirs was studied
in\cite{BSJP:2021,AMIS:2022,BSPEPANL:2022}, where it was shown
that the spectrum of the low-frequency fluorescence, as well as
the propensity of the quantum system in question to either absorb
or amplify low-frequency radiation, can be effectively altered by
controlling the parameters of the squeezed dissipative reservoir.
These results lead to a reasonable assumption that affecting a
polar quantum system, driven by external nearly resonant
high-frequency monochromatic fields, by one more properly attuned
low-frequency monochromatic field,  its frequency being within the
range of the Rabi frequencies of the exciting high-frequency
fields, one can alter and control the spectral properties of the
resonance fluorescence in the high-frequency range. To check this
assumption, we consider, hereinafter,  a polar two-level system
driven by an arbitrary polychromatic field in general, with a
particular attention given to the case of bichromatic and
trichromatic excitation.

\section{ Two-Level Polar Quantum System in a Polychromatic Field}

Despite the fact that real physical systems employed in quantum
optical studies are mind-boggling complex, even simple theoretical
models may provide one with sufficient understanding of the
physical mechanisms of optical phenomena observed in experiments,
as well as to predict new effects useful for practical
applications. Among them, a broad class of various two-level
models, characterized by that the quantum system in question is
reduced to only two energy levels, came to outstanding prominence
in theoretical studies of resonance fluorescence. The optical
properties of these models are determined by spatial symmetries,
revealed quantitatively in terms of multipolar transition moments.
 In what follows, we are considering a polar two-level "atom" placed
at the atomic position $\vr_0$ with the ground state
$|\,0\rangle$, excited state $|1\rangle$ and transition frequency
$\om_0$, driven by external stationary polychromatic field ${\bf
E}_{ext}(t)=\dsum_{j=1}^N{\bf E}^{(j)}\cos(\om_j t+\phi_j)$ with
the amplitudes ${\bf E}^{(j)}$, frequencies $\om_j$ and phases
$\phi_j$. Simultaneously, this two-level system interacts with a
bosonic dissipative reservoir $F$ comprised of the vacuum modes of
quantized electromagnetic field.  Therefore, the system
Hamiltonian  takes the form of the generalized Rabi model
\cite{RabiC,Scully}

\bg H=H_0(S)+H_0(F)+H_{SF}+H_{SE}(t),\label{model}
 \en

\bg   \hspace{-2.9cm}\mbox{where}\hspace{2.0cm} H_0(S)=
\hb\om_0S^z, \quad H_0(F)= \dsum_{\vk,\lam}\hb\nu_{\vk}\akla\akl,
\en

\bg\quad H_{SE}(t)=-e\dsum_{j=1}^N\hat{\bf r}\cdot{\bf
E}^{(j)}\cos(\om_j t+\phi_j)=-\dsum_{j=1}^3 \hat{\bf d}\cdot{\bf
E}^{(j)}\cos(\om_j t+\phi_j), \en

\bg  H_{SF}=\hb\dsum_{l,j}\dsum_{\vk,\lam}
g_{\vk,\lam}^{lj}\sigma_{lj}(\akla e^{-i\vk\vr_0}+\akl
e^{i\vk\vr_0}),\quad l =0,1, \quad j=0,1, \en

\noi  where the operator of the atomic  electric dipole moment
${\hat{\bf d}}=e\hat\vr$ is defined as

\bg \hat{\bf d}=\dsum_{l,j}{\bf d}_{lj}\sigma_{lj},\, {\bf
d}_{lj}=e\lan l|\hat\vr |j\ran,\, \sigma_{lj}= |l\ran\lan
j|,\,[\sigma_{lj},\sigma_{kp}]=\sigma_{lp}\delta_{jk}-\sigma_{kj}\delta_{lp},\en

\noi
 $b_{\vk,\lam}$ and $b^+_{\vk,\lam}$ are the bosonic
operators of the creation and annihilation of photons with the
wave vector $\vk$ and the polarization  $\lam$, which obey the
following commutation relations

\bg [b_{\vk,\lam},b^+_{\vk',\lam'}]=
\delta_{\vk,\vk'}\delta_{\lam,\lam'},\quad
[b^+_{\vk,\lam},b^+_{\vk',\lam'}]= 0,\quad
[b_{\vk,\lam},b_{\vk',\lam'}]=0, \en

\bg \qquad g_{\vk,\lam}^{lj}=-\fr{{\bf d}_{lj}\cdot \vekl {\cal
E}_{\vk}}{\hb},\quad {\cal
E}_\vk=\left(\fr{\hb\nu_\vk}{2\veps_0V}\right)^{1/2}, \nu_\vk =
c|\vk|,   \en

\bg \dsum_{\vk,\lam}\to
\dsum_{\lam}\left(\fr{L}{2\pi}\right)^3d^3k,\, k_l=\fr{2\pi
n_l}{L},\, V=L^3,\,\, n_l=0\pm 1,\pm 2,...,\, l=x,y,z,
 \en

\noi the polarization vectors $\vekl$ of the quantized field
satisfy the relationships

\bg \vk\cdot\vekl=0,\quad \veps^{(1)}_{\vk l}\veps^{(1)}_{\vk j}
+\veps^{(2)}_{\vk l}\veps^{(2)}_{\vk
j}=\delta_{lj}-\fr{k_lk_j}{k^2}, \quad l,j=x,y,z,\en

\noi  and $S^z=\fr{1}{2}(|1\rangle\langle 1| - |\,0\rangle\langle
0|)$ is the inverse population operator. The term $H_{SE}(t)$
describing the interaction with external field is given by

\bg H_{SE}(t)=\dsum^N_{j=1}\fr{\hb}{2}\left(e^{i(\om_j
t+\phi_j)}+e^{-i(\om_j t+\phi_j}\right)\times \nn\en

\bg\times \left[\Om_R^{\,(j)}( S^++ S^-) + \delta_a^{(j)} S^z -
\fr{\delta_s^{(j)}}{2} (|1\rangle\langle 1|+ |0\rangle\langle
0|)\right], \label{sf1}
 \en

\noi where  $S^+=|1\rangle\langle 0|$ and $S^-=|\,0\rangle\langle
1|$ are the raising and lowering atomic operators, $
\Om_R^{(j)}=-{\bf E}^{(j)} \cdot{\bf d}_{10}/\hb$ are the Rabi
frequencies of the field components, which, without loss of
generality, can be assumed to be real and positive, and

\vspace{-0.1cm}

\bg {\bf d}_{10}=e\langle 1|\hat {\bf r}|0\rangle,\quad {\bf
d}_{01}=e\langle 0|\hat {\bf r}|1\rangle,\quad {\bf
d}_{11}=e\langle 1|\hat {\bf r}|1\rangle,\quad {\bf
d}_{00}=e\langle 0|\hat {\bf r}|0\rangle\en

\noi are the matrix elements of the atomic  electric dipole
operator.  As was already mentioned above, we neglect the
assumption ${\bf d}_{11}={\bf d}_{00}=0$ in favor of a more
general assumption ${\bf d}_{11}\ne{\bf d}_{00}$, being a
consequence of the broken inversion symmetry. Therefore,

\bg\delta_s^{(j)}={\bf E}^{(j)}\cdot({\bf d}_{00}+{\bf
 d}_{11})/\hb,  \qquad \delta_a^{(j)}={\bf E}^{(j)}\cdot({\bf d}_{00}-{\bf
 d}_{11})/\hb\en

 \noi are the  symmetry violation parameters, and besides, the
 terms  proportional to  $\delta_s^{(j)}$ do not influence the dynamics of the system
 and can be neglected.

\section{The Fluorescence Spectrum}

\noi The incoherent part of the steady-state fluorescence spectrum
is given, as usual \cite{Scully}, by

\bg \hspace{-2.5cm} S_{in}(\om)\sim
\fr{\Gamma}{\pi}\mbox{Re}\left[ \lim_{T\to\infty}\fr{1}{T}
\dint_0^T dt\dint_0^{T-t} d\tau[\lan S^+(t)S^-(t+\tau)\ran -  \lan
S^+(t)\ran\lan S^-(t+\tau)\ran
]e^{i\om\tau}\right].\label{fspecin}\en

\noi  Temporal evolution of the "atomic" subsystem $S$ is
determined by the master equation for the reduced statistical
operator $\rho_S(t)$ being derived by conventional methods
\cite{Puri:2001,Carm} in the form

\bg \hspace{-2.0cm}\fr{\partial \rho_S(t)}{\partial
t}=-\fr{i}{\hb}[H_{SE}(t),\rho_S(t)]-\fr{1}{2}\Gamma
(S^+S^-\rho_S(t)+\rho_S(t)S^+S^--2S^-\rho_S(t)S^+),\label{master}\en

\noi where $\Gamma$ is the radiative damping constant due to the
atomic interaction with a bosonic dissipative environment.
 In what follows, let us concentrate our attention
on the case of the commensurable frequencies $\om_j$, $j=1,...,N$,
and the average frequency  $\om_s=\dsum_{j=2}^N\om_j/(N-1)$ of the
external driving fields ${\bf E}^{(2)},...,{\bf E}^{(N)}$, and
introduce the fundamental frequency $\nu=\om_s/p=\om_j/n_j$, with
$n_j, p$ being some arbitrary integers.  Then, an infinite set of
equations for the slowly varying amplitudes $X_i^{(l)}(t)$ of the
averaged atomic variables of interest $\lan \tilde S^\mp(t)\ran$
and $\lan \tilde S^z(t)\ran$ follows readily from
Eq.(\ref{master}):

\ba \fr{d}{dt} X_1^{(l)}(t)=-\left(\fr{\Gam}{2}+ i\Delta+il\nu
\right) X_1^{(l)}(t) -\nonumber\ea \bg
-i\dsum_{j=1}^N\fr{\delta_a^{(j)}}{2}\left(e^{i\phi_j}X_1^{(l-n_j)}(t)+e^{-i\phi_j}X_1^{(l+n_j)}(t)\right)
+\nn\en

\bg
+\dsum_{j=1}^N\Om^{(j)}_R\left(e^{i\phi_j}X_3^{(l-(n_j+p))}(t)+e^{-i\phi_j}X_3^{(l-(p-n_j))}(t)\right),
\label{emx1}\en

\ba \fr{d}{dt} X_2^{(l)}(t)=-\left(\fr{\Gam}{2}- i\Delta+il\nu
\right) X_2^{(l)}(t) + \nn\ea

\bg
+i\dsum_{j=1}^N\fr{\delta_a^{(j)}}{2}\left(e^{i\phi_j}X_2^{(l-n_j)}(t)+e^{-i\phi_j}X_2^{(l+n_j)}(t)
\right) + \nn \en

\bg+\dsum_{j=1}^N\Om_R^{(j)}
\left(e^{-i\phi_j}X_3^{(l+(n_j+p))}(t)+e^{i\phi_j}X_3^{(l+(p-n_j))}(t)\right),
\label{emx2}\en

\bg \fr{d}{dt} X_3^{(l)}(t)
=-\fr{\Gam}{2}\delta_{l,0}-(\Gam+il\nu)X_3^{(l)}(t)-\nn\en

\bg
 -\dsum_{j=1}^N\fr{\Om_R^{(j)}}{2}\left(e^{i\phi_j}X_1^{(l-(n_j-p))}(t)+e^{-i\phi_j}X_1^{(l+(n_j+p))}(t)
 + \right.\nn\en

\bg\left.+
e^{i\phi_j}X_2^{(l-(n_j+p))}(t)+e^{-i\phi_j}X_2^{(l+(n_j-p))}(t)
 \right),\label{emx3}\en

\noi where

\bg  \tilde S^-(t)  = -i e^{ i\om_s t }S^-(t),\quad \tilde S^+(t)
= i e^{ -i\om_s t } S^+(t),\quad \tilde S^z(t)=S^z(t), \en

\bg \lan \tilde S^-(t) \ran =- i e^{ i\om_s t } \mbox{Sp}(S^-
\rho_S(t))= \dsum_{l=-\infty}^{+\infty}X_1^{(l)}(t)e^{il\nu t},
\label{decomp1} \en

\bg \lan \tilde S^+(t) \ran = i e^{- i\om_s t } \mbox{Sp}(S^+
\rho_S(t))= \dsum_{l=-\infty}^{+\infty}X_2^{(l)}(t)e^{il\nu t},
\label{decomp2} \en

\bg \lan \tilde S^z(t) \ran = \mbox{Sp}(S^z \rho_S(t))=
\dsum_{l=-\infty}^{+\infty}X_3^{(l)}(t)e^{il\nu t},
\label{decomp3}\en

\noi and $\Delta=\om_0-\om_s$ is the mismatch between  the atomic
transition frequency and the average frequency $\om_s$.

\noi Because of the  quantum regression theorem, the correlation
functions \ba Y_1(t,t+\tau)=\lan \tilde S^+(t)\tilde
S^-(t+\tau)\ran-\lan \tilde S^+(t)\ran\lan\tilde
S^-(t+\tau)\ran,\nn\ea

\ba Y_2(t,t+\tau)=\lan \tilde S^+(t)\tilde S^+(t+\tau)\ran-\lan
\tilde S^+(t)\ran \lan \tilde S^+(t+\tau)\ran,\nn\ea

\ba Y_3(t,t+\tau)=\lan \tilde S^+(t)\tilde S^z(t+\tau)\ran -\lan
\tilde S^+(t)\ran\lan\tilde S^z(t+\tau)\ran, \nn\ea

\noi  obey nearly the same set of the temporal evolution equations
as the one for the correspondent averaged atomic variables
$\lan\tilde S^-(\tau)\ran$, $\lan\tilde S^+(\tau)\ran$ and
$\lan\tilde S^z(\tau)\ran$. The only difference is that  their
right-hand sides do not contain the inhomogeneous term $-\Gamma
/2$  as a result of the subtraction of the coherent contribution
terms $\lan \tilde S^+(t)\ran\lan\tilde S^-(t+\tau)\ran$,  $\lan
\tilde S^+(t)\ran\lan\tilde S^+(t+\tau)\ran$ and $\lan \tilde
S^+(t)\ran\lan\tilde S^z(t+\tau)\ran$. As was already done above
for the averaged atomic variables, these correlation functions can
also be decomposed as

\bg
Y_i(t,t+\tau)=\dsum_{l=-\infty}^{+\infty}Y_i^{(l)}(t,\tau)e^{il\nu(t+
\tau)},\quad i=1,2,3, \label{decomp4} \en

\noi and the Laplace transforms

\bg \bar Y_i^{(l)}(t,z)=\dint_0^\infty
e^{-z\tau}Y_i^{(l)}(t,t+\tau)d\tau \label{LaplY} \en \noi

\noi of their slowly varying amplitudes $Y_i^{(l)}(t,\tau),\,
i=1,2,3$, will satisfy the following set of equations:

\ba z\bar Y_1^{(l)}(t,z)+\left(\fr{\Gam}{2}+ i\Delta+il\nu \right)
\bar Y_1^{(l)}(t,z) +\nonumber\ea

\bg +i\dsum_{j=1}^N\fr{\delta_a^{(j)}}{2}\left(e^{i\phi_j}\bar
Y_1^{(l-n_j)}(t,z)+e^{-i\phi_j}\bar Y_1^{(l+n_j)}(t,z)\right)
-\nn\en

\bg -\dsum_{j=1}^N\Om^{(j)}_R\left(e^{i\phi_j}\bar
Y_3^{(l-(n_j+p))}(t,z)+e^{-i\phi_j}\bar
Y_3^{(l-(p-n_j))}(t,z)\right)= \en

\ba
=\fr{1}{2}\delta_{l,0}+X_3^{(l)}(t)-\dsum_{r=-\infty}^{\infty}X_1^{(l-r)}(t)X_2^{(r)}(t),
\label{emyz1}\ea

\ba  z\bar Y_2^{(l)}(t,z) +\left(\fr{\Gam}{2}- i\Delta+il\nu
\right)  \bar Y_2^{(l)}(t,z)  - \nn\ea

\bg -i\dsum_{j=1}^N\fr{\delta_a^{(j)}}{2}\left(e^{i\phi_j}\bar
Y_2^{(l-n_j)}(t,z)+e^{-i\phi_j}\bar Y_2^{(l+n_j)}(t,z) \right)
 - \nn \en

\bg-\dsum_{j=1}^N\Om_R^{(j)} \left(e^{-i\phi_j}\bar
Y_3^{(l+(n_j+p))}(t,z)+e^{i\phi_j}\bar
Y_3^{(l+(p-n_j))}(t,z)\right)=\nn \en

\ba=
-\dsum_{r=-\infty}^{\infty}X_2^{(l-r)}(t)X_2^{(r)}(t)\label{emyz2}\ea

\bg z\bar Y_3^{(l)}(t,z)+(\Gam+il\nu)
  \bar Y_3^{(l)}(t,z) +\nn\en

\bg
 +\dsum_{j=1}^N\fr{\Om_R^{(j)}}{2}\left(e^{i\phi_j}\bar Y_1^{(l-(n_j-p))}(t,z)
 +e^{-i\phi_j}\bar Y_1^{(l+(n_j+p))}(t,z) +\right.\nn \en

 \bg \left.+ e^{i\phi_j}\bar Y_2^{(l-(n_j+p)}(t,z)+e^{-i\phi_j}\bar Y_2^{(l+(n_j-p))}(t,z)
 \right)=\nn\en

\ba= -\dsum_{r=-\infty}^{\infty}\left(
\fr{1}{2}\delta_{r,0}+X_3^{(r)}(t)  \right)X_2^{(l-r)}(t).
\label{emyz3}\ea

\noi  In the steady state limit $(T\to\infty)$ only the zero-order
component $\bar Y_1^{(0)}(t,z)$ contributes to $S_{in}(\om)$, and
the incoherent part of the spectrum  reads finally as (cf.
\cite{FicekSwain:2005}):

\bg \!\!S_{inc}(\om)\sim
\fr{\Gam}{\pi}\mbox{Re}\lim_{T\to\infty}\fr{1}{T}\dint_0^T\!dt\dint_0^{T-t}
d\tau Y_1(t,t+\tau) e^{i(\om-\om_s)\tau} = \nn\en

\bg =\fr{\Gam}{\pi}\mbox{Re} \lim_{t\to\infty}\bar
Y_1^{(0)}(t,z)\Big |_{z=-i(\om-\om_s)}.\label{fspecinY0}\en

\section{Numerical Results}

\begin{figure}[th]
\vspace{1.0cm}\hspace{2pc}
\begin{minipage}{22.5pc}
\includegraphics[width=22.5pc]{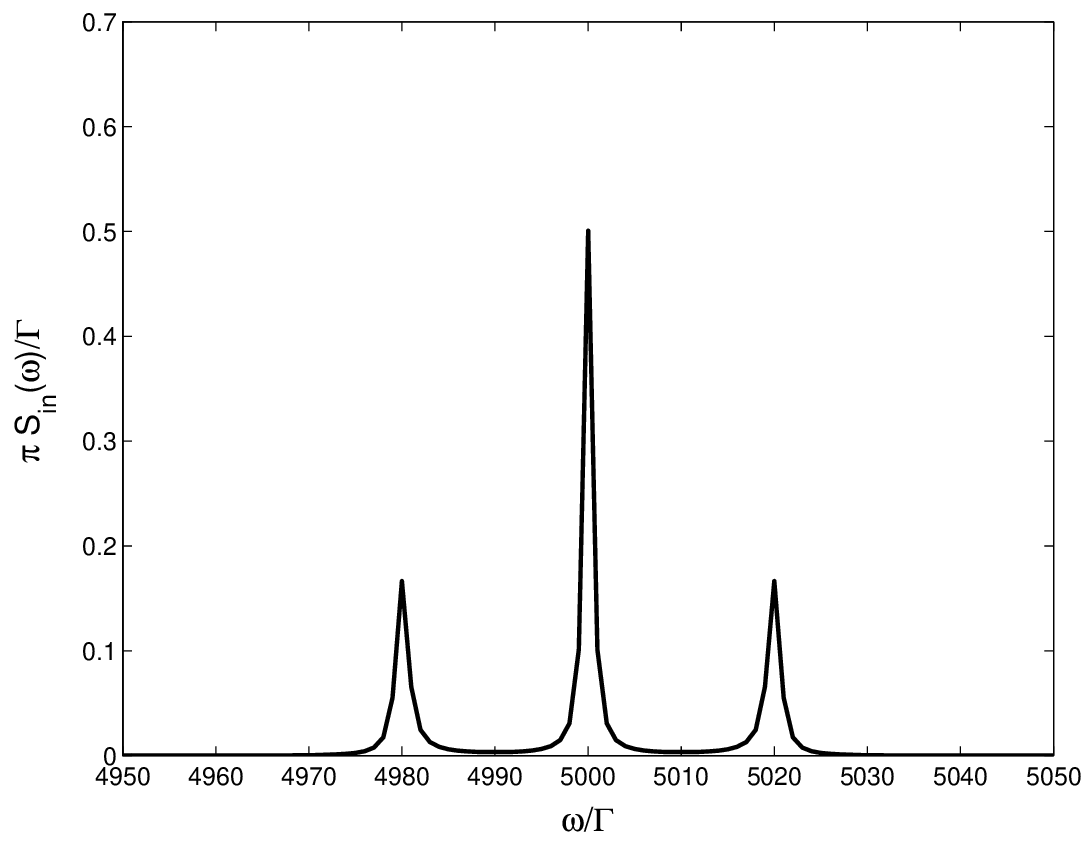} \caption{
Fluorescent Mollow triplet for a non-polar two-level quantum
system. \, $\Gamma=1,\,\, \om_1=\om_0=5000\Gamma,  \,\,
\Om_R^{(1)}=20\Gam.$ }\label{f1tor}
\end{minipage}
\end{figure}

\begin{figure}[th]
\vspace{0.0cm}\hspace{2pc}
\begin{minipage}{22.5pc}
\includegraphics[width=22.5pc]{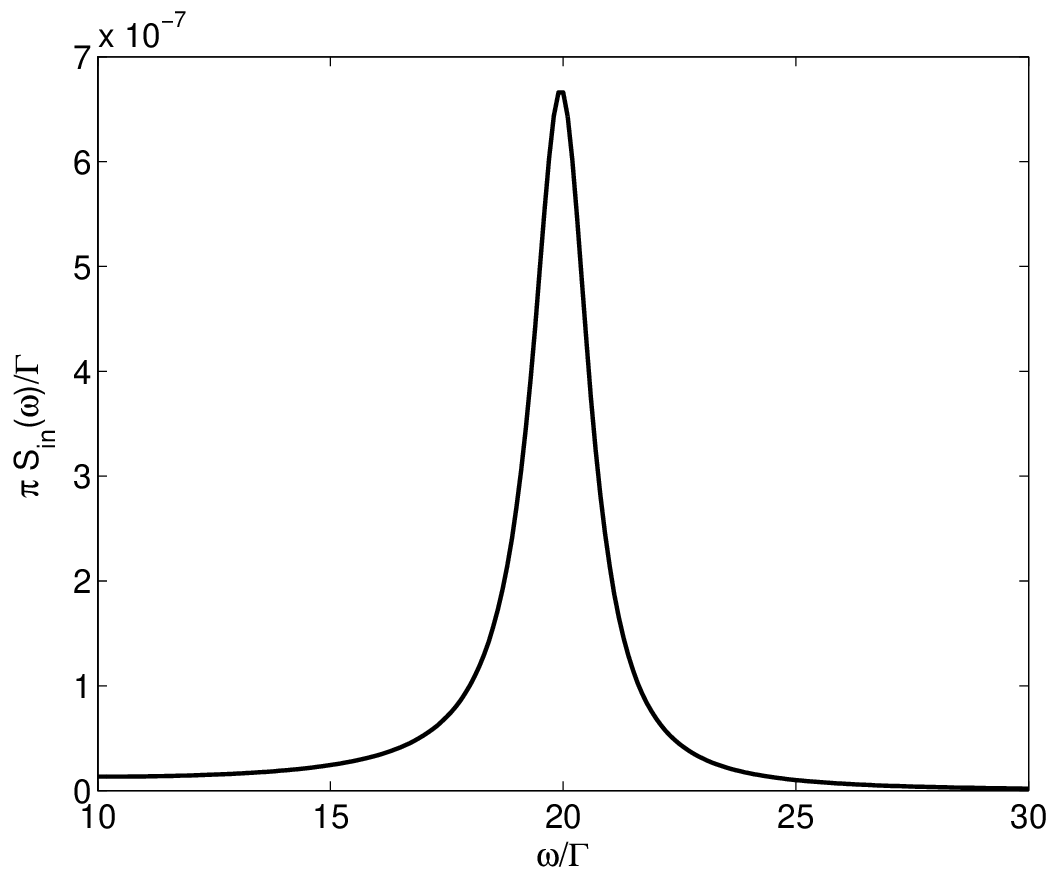}
\caption{Spectral peak near the frequency $\om=\Om_R^{(1)}$ in the
case  $\Gamma=1, \om_1=\om_0=5000\Gam,
\Om_R^{(1)}=\delta_a^{(1)}=20\Gam,
\Om_R^{(2)}=\delta_a^{(2)}=0.$}\label{f3tor}
\end{minipage}
\end{figure}

\begin{figure}[th]
\hspace{2pc}
\begin{minipage}{22.5pc}\vspace{-0.6pc}
\includegraphics[width=22.5pc]{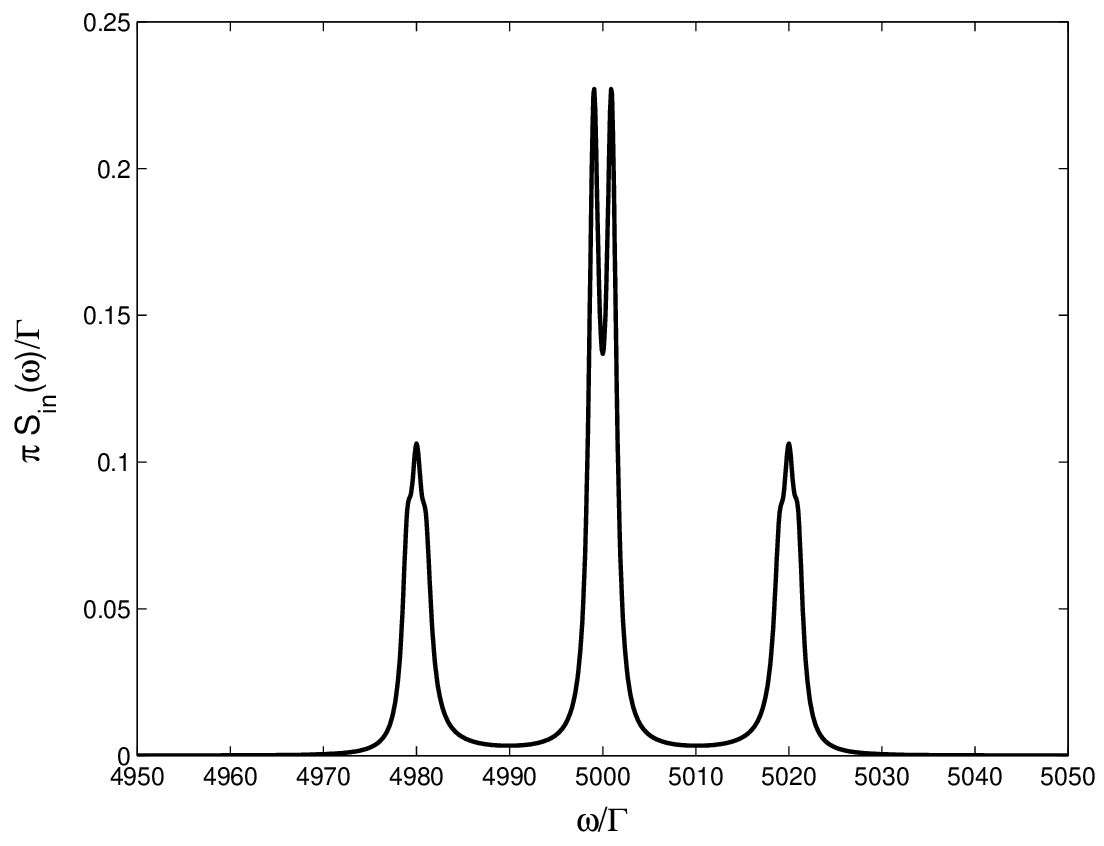}\vspace{-0.5pc}
\caption{Splitting of the central peak of the Mollow triplet in
the case $\Gamma = 1, \om_1=\om_0=5000\Gam, \om_2=\Om_R^{(1)} =
20\Gam, \delta_a^{(1)}=20\Gam$, and $\Om_R^{(2)} = \delta_a^{(2)}
= 2\Gam$.}\label{f4tor}
\end{minipage}
\end{figure}

\begin{figure}[th]
\vspace{0.0cm}\hspace{2pc}
\begin{minipage}{22.5pc}
\includegraphics[width=22.5pc]{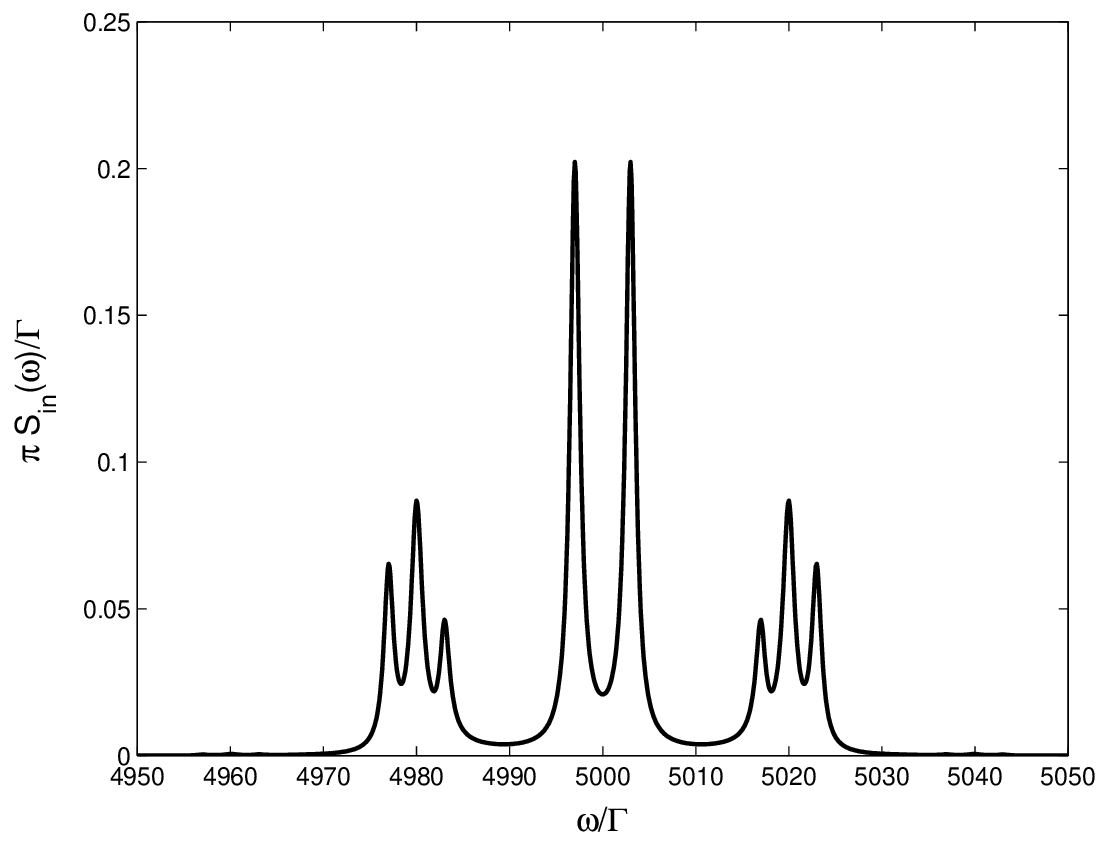}
\caption{Transformation of the Mollow triplet into a spectrum
consisting of a spectral doublet and two triplets in the case
$\Gamma = 1, \om_1 = \om_0 = 5000\Gam, \om_2=\Om_R^{(1)} =
\delta_a^{(1)} = 20\Gam$, and $\Om_R^{(2)} = \delta_a^{(2)} =
6\Gam$.} \label{f5tor}
\end{minipage}
\end{figure}

\begin{figure}[th]
\vspace{0.5cm}\hspace{2pc}
\begin{minipage}{22.5pc}\vspace{-0.6pc}
\includegraphics[width=22.5pc]{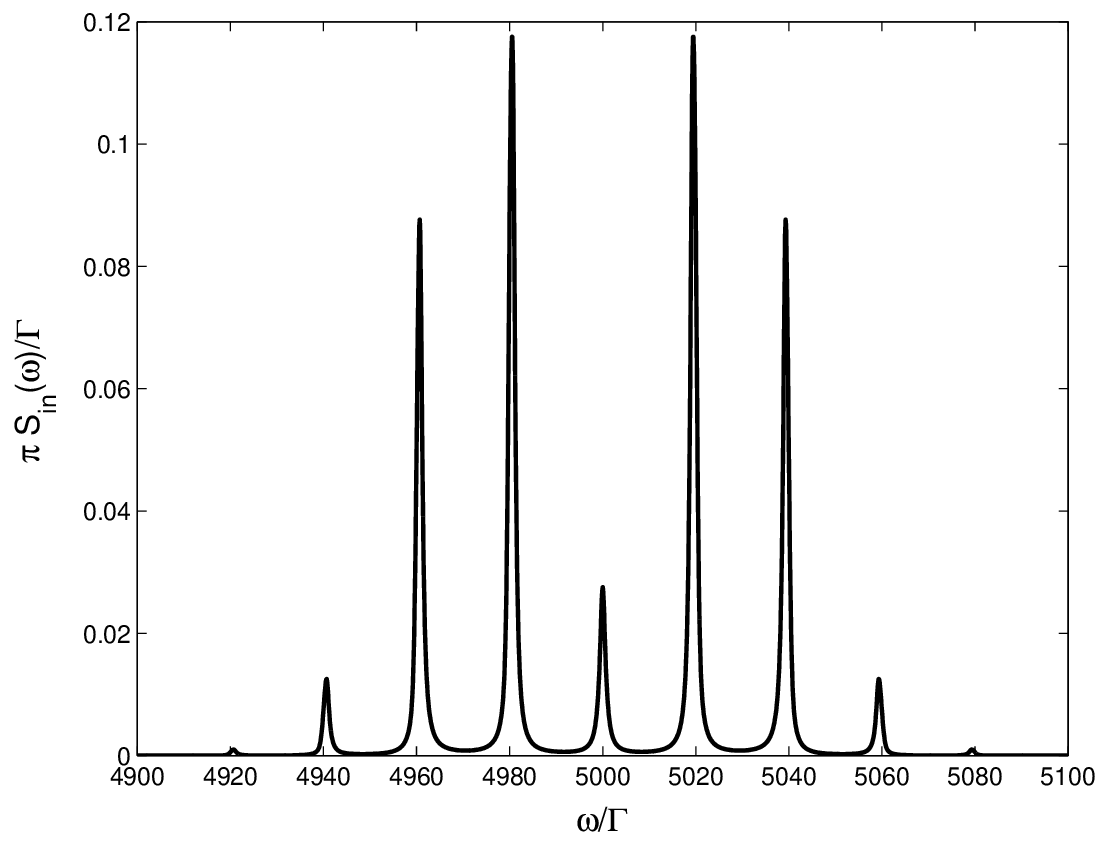}\vspace{-0.5pc}
\caption{Equidistant fluorescence spectrum in the case $\Gamma =
1, \om_1 = \om_0 = 5000\Gam, \om_2=\Om_R^{(1)} = \delta_a^{(1)} =
20\Gam$, and $\Om_R^{(2)} = \delta_a^{(2)} = 40\Gam.$}
\label{f6tor}
\end{minipage}
\end{figure}

\begin{figure}[th]
\vspace{0.0cm}\hspace{2pc}
\begin{minipage}{22.5pc}
\includegraphics[width=22.5pc]{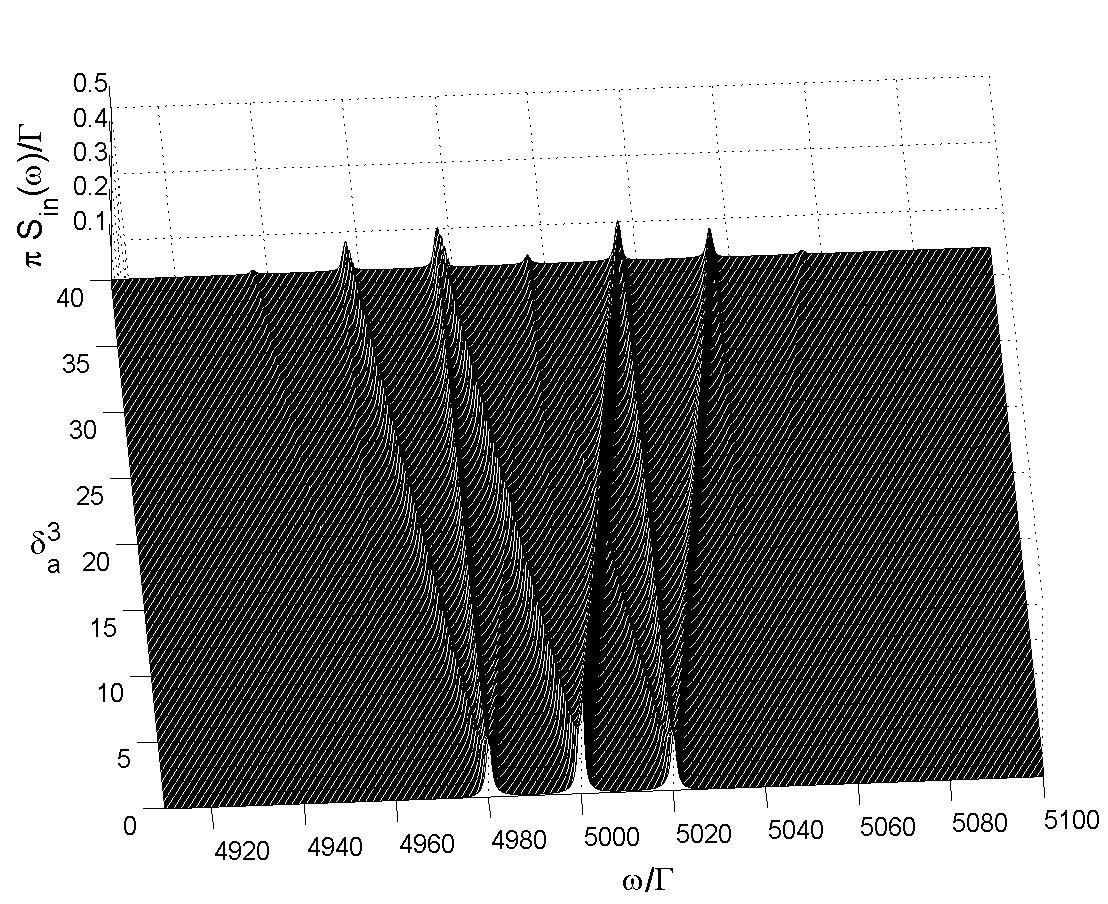}
\caption{Complete evolution of the high-frequency fluorescent
spectrum depending on the asymmetry parameter $\delta_a^{(2)}$.
$\Gamma = 1, \om_1 = \om_0 = 5000\Gam, \om_2=\Om_R^{(1)} =
\delta_a^{(1)} = 20\Gam$, and $\Om_R^{(2)} = \delta_a^{(2)} =
(0:1:40)\Gam.$} \label{f7tor}
\end{minipage}
\end{figure}

As usual, Eqs.(\ref{emx1})-(\ref{emx3}) and
Eqs.(\ref{emyz1})-(\ref{emyz3})  were analyzed numerically in the
steady-state limit $(t\to\infty)$ by restricting the number of
amplitudes $X_i^{(l)}(t)$ and $\bar Y_i^{(l)}(t,z)$ taken into
account in the calculations. Two cases were studied in total. In
the first one, the polar system was excited by a bichromatic field
consisting of a high-frequency resonance component with the
frequency coinciding with the atomic transition frequency, and a
low-frequency component whose frequency coincides with the Rabi
frequency of the high-frequency component. In the second case, the
system was driven by a trichromatic  field comprized of two
nearly-resonant high-frequency bichromatic field components of
equal Rabi frequencies and their carrier frequencies being placed
symmetrically around the atomic transition frequency, and one more
low-frequency component being in resonance with the Rabi frequency
of the high-frequency components. In the bichromatic case
considered, the external exciting field frequency $\om_1$ is in
resonance with the atomic transition frequency $\om_0$. In the
absence of the low-frequency field ${\bf E}^{(2)}$, the
high-frequency spectrum has a characteristic shape (see
Fig.\ref{f1tor}) known as the Mollow triplet \cite{Mollow1}, but
in the low-frequency part, there is a spectral peak (see
Fig.\ref{f3tor})
 located almost at the Rabi frequency $\Om_R^{(1)}$ of the exciting field ${\bf
E}^{(1)}$, which is typical of polar two-level quantum systems
excited by a resonant monochromatic field
\cite{Kibis:2009,Soldatov:2016}. In this case, the polarity
property of the quantum system has a negligible effect on the
shape of the high-frequency spectrum. Then, let us consider a very
realistic experimental situation, with the same polar two-level
system simultaneously subjected to the effect of the resonant
($\om_2 = \Om_R^{(1)})$ low-frequency field of increasing
intensity, such that the relation $\delta_a^{(2)}/\Om_R^{(2)} =
\mbox{const}$ is naturally satisfied due to the definitions of
these quantities. As it turned out, for a nonpolar quantum system
with $\delta_a^{(1)}=\delta_a^{(2)}=0$, the shape of the
high-frequency spectrum in Fig.\ref{f1tor} is independent of the
effect of the low-frequency component of the bichromatic field
with the frequency $\om_2=\Om_R^{(1)}$ and coincides with the
spectrum predicted in \cite{Mollow1}, even in the case where the
intensity of the low-frequency component is an order of magnitude
higher than the intensity of the high-frequency component, that
is, when the condition $\Om_R^{(1)} <<\Om_R^{(2)} << \om_0$ is
satisfied. Contrariwise, as can be seen from Fig.\ref{f4tor} , in
the case of a polar quantum system, the splitting of the central
peak of the Mollow triplet is observed even if the intensity of
the low-frequency component is an order of magnitude lower than
the intensity of the high-frequency component, $\Om_R^{(2)} <<
\Om_R^{(1)}$. The side peaks of the triplet generally preserve
their shape except for some narrowing of their peaks, a decrease
in amplitudes, and a slight broadening, which become more
pronounced as the intensity of the low-frequency component
increases. At the same time, an ever deeper splitting of the
central peak occurs, which ultimately leads to the almost complete
disappearance of fluorescence at the central frequency of the
atomic transition $\om=\om_0$ and the transformation of the
central peak of the fluorescence spectrum into a doublet whose
components are centered at the frequencies $\om_0 \pm
\delta_a^{(2)}/2$ (see Fig.\ref{f4tor}). In this case, the side
peaks of the Mollow triplet transform into spectral triplets that
resemble the original Mollow triplet, but are centered at the
frequencies $\om_0 \pm \Om_R^{(1)}$. At the same time, as can be
seen from Fig.\ref{f5tor}, the side components of these triplets
are located at the frequencies $\om_0 + \Om_R^{(1)} \pm
\delta_a^{(2)}/2$ and $\om_0 - \Om_R^{(1)} \pm \delta_a^{(2)}/2$,
that is, the quantity $\delta_a^{(2)}/2$ plays the same role in
the formation of side triplets as the Rabi frequency $\Om_R$ does
in the formation of the initial Mollow triplet in Fig.\ref{f1tor}
in the case of a monochromatic exciting field. The situation in
which the low-frequency component of the bichromatic field is more
intense than the high-frequency component, that is, $\Om_R^{(2)} >
\Om_R^{(1)}$, also leads to interesting and potentially useful
results, including the formation of a high-frequency equidistant
fluorescence spectrum with good spectral resolution at a certain
relation for the intensities of the bichromatic field components
and the parameters of the quantum model. This spectrum consists of
the restored central peak at the atomic transition frequency
surrounded by four doublets symmetric with respect to this
frequency, and all peaks in this structure are spaced from each
other at a distance equal to $\delta_a^{(2)}/2$ (see
Fig.\ref{f6tor}). Complete evolution of the fluorescent spectrum
depending on the asymmetry parameter $\delta_a^{(2)}$ can be seen
in Fig.\ref{f7tor}. It is worth to note, that all the effects
described above are independent of the relative phase
$\phi_2-\phi_1$, which, of course, should facilitate their
experimental research and practical application.

\begin{figure}[th]
\hspace{2pc}
\begin{minipage}{22.5pc}\vspace{-0.6pc}
\includegraphics[width=22.5pc]{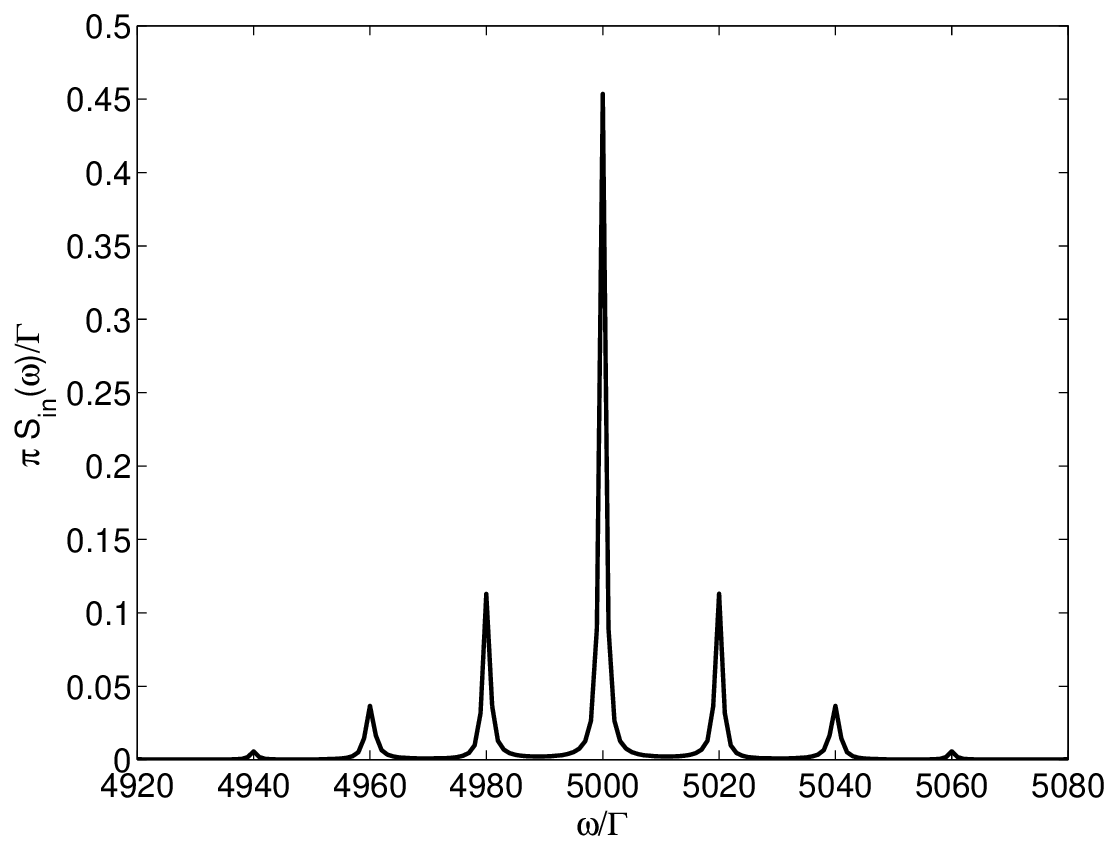}\vspace{-0.5pc}
\caption{ High-frequency fluorescent spectrum for a non-polar
trichromatically driven quantum system. $\Gamma=1,
\om_0=\om_s=5000\Gam, \delta=20\Gam, \om_1=\om_0-\delta,
\om_2=\om_0+\delta, \Om_R^{(1)}=\Om_R^{(2)}=\Om_R=20\Gam,
\om_3=\Om_R, \Om_R^{(3)}=20\Gam,
\delta_a^{(1)}=\delta_a^{(2)}=\delta_a^{(3)}=0, \phi_j=0,
j=1,2,3.$} \label{f8tor}
\end{minipage}
\end{figure}

\begin{figure}[th]
\vspace{0.0cm}\hspace{2pc}
\begin{minipage}{22.5pc}
\includegraphics[width=22.5pc]{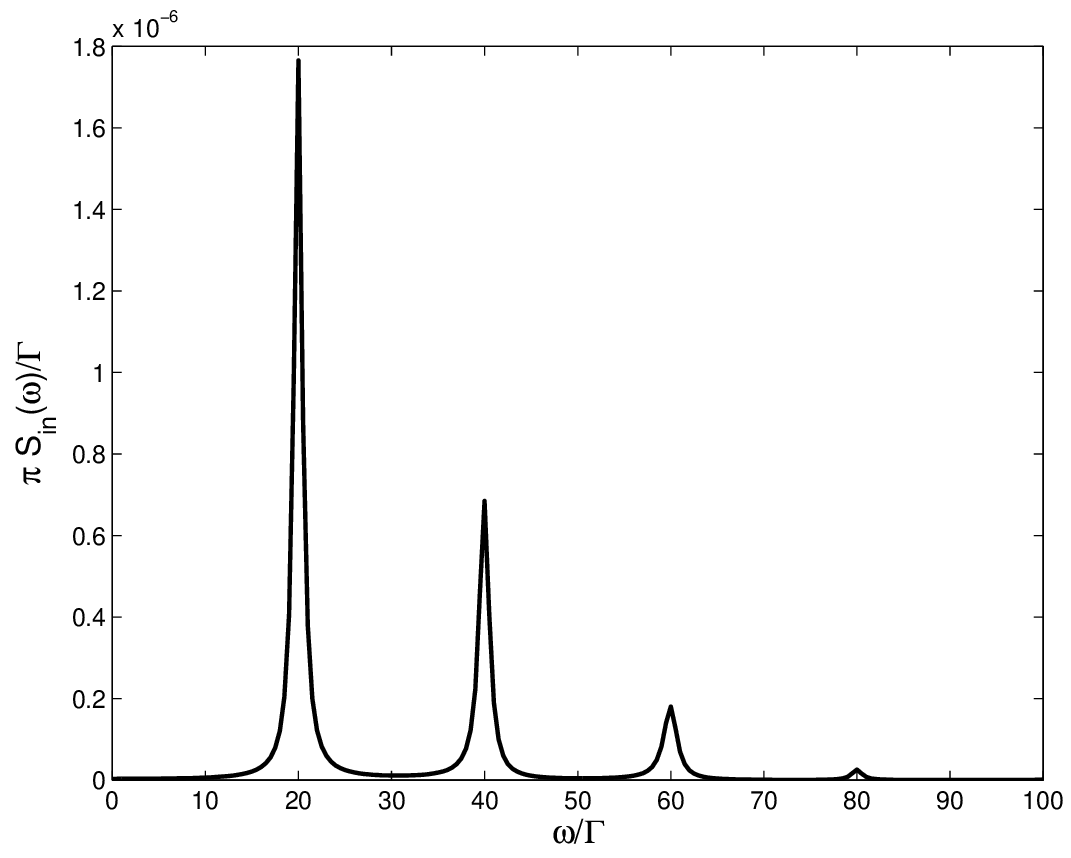}
\caption{Low-frequency fluorescent spectrum for a polar
bichromatically driven quantum system. $\Gamma=1,
\om_0=\om_s=5000\Gam, \delta=20\Gam, \om_1=\om_0-\delta,
\om_2=\om_0+\delta, \Om_R^{(1)}=\Om_R^{(2)}=\Om_R=20, \Om_R^3=0,
\delta_a^{(1)}=\delta_a^{(2)}=20\Gam, \delta_a^{(3)}=0,  \phi_j=0,
j=1,2,3.$} \label{f9tor}
\end{minipage}
\end{figure}

\begin{figure}[th]
\hspace{2pc}
\begin{minipage}{22.5pc}\vspace{-0.6pc}
\includegraphics[width=22.5pc]{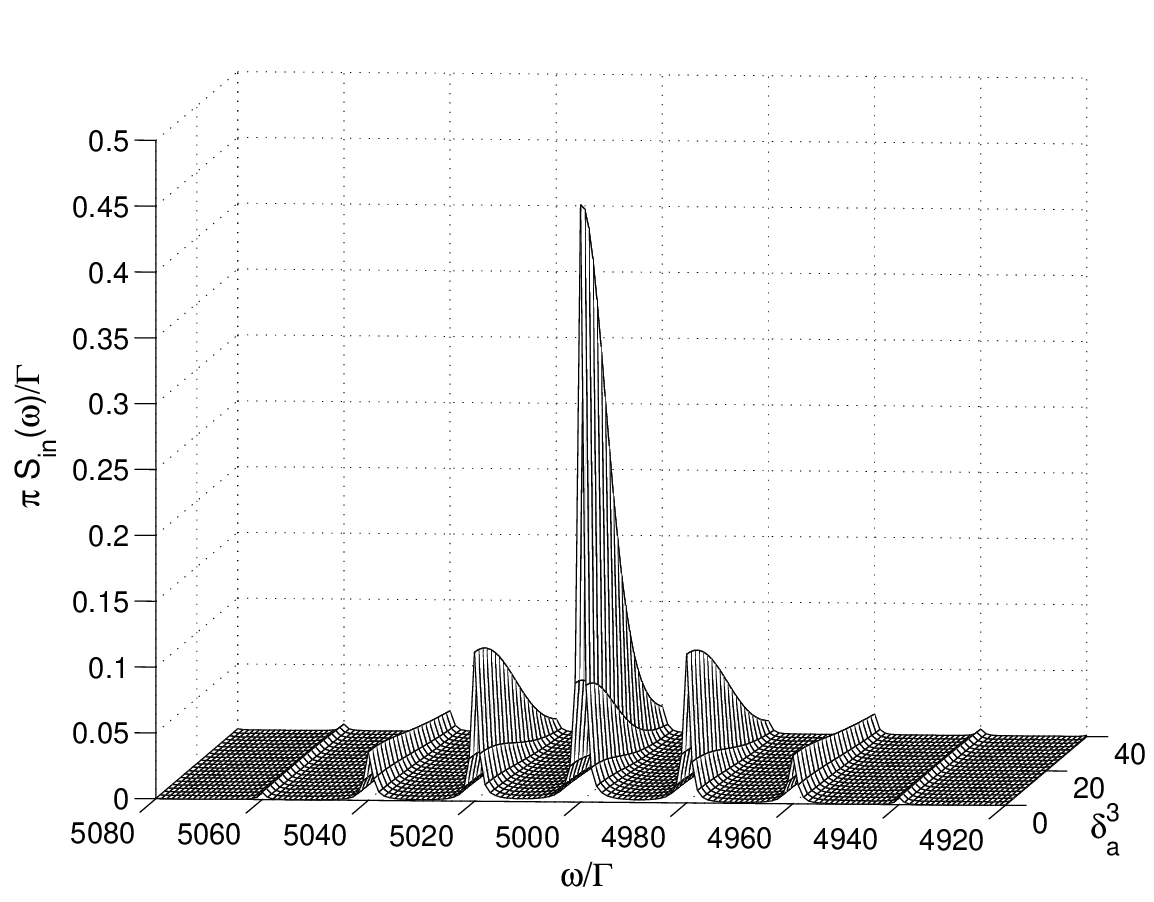}\vspace{-0.5pc}
\caption{ Complete evolution of the high-frequency fluorescent
spectrum depending on the asymmetry parameter $\delta_a^3$ for a
polar trichromatically driven quantum system. $\Gamma=1,
\om_0=\om_s=5000\Gam, \delta=20\Gam, \om_1=\om_0-\delta,
\om_2=\om_0+\delta, \om_3=\Om_R,
\Om_R^{(1)}=\Om_R^{(2)}=\Om_R=20\Gam, \Om_R^{(3)}=20,
\delta_a^{(1)}=\delta_a^{(2)}=20\Gam, \delta_a^{(3)}=(0:2:40)\Gam,
\phi_j=0, j=1,2,3.$}\label{f10tor}
\end{minipage}
\end{figure}

 In the case of
the trichromatic excitation a polar two-level system is driven by
an external field, with two high-frequency bichromatic field
components of equal Rabi frequencies
$\Om_R^{(1)}=\Om_R^{(2)}=\Om_R$ and carrier frequencies
$\om_1=\om_0-\delta, \om_2=\om_0+\delta$ detuned symmetrically
from the atomic transition frequency $\om_0$, and one more
low-frequency component with frequency $\om_3$ being in resonance
with the Rabi frequency $\Om_R$ of the high-frequency components.
It was found that  the high-frequency fluorescence spectrum of the
non-polar system driven this way contains well-separated peaks at
frequencies $\om_0\pm n\delta, n=0,1,2,...,$ provided that the
detuning $\delta$ is much greater than the width of the excited
level $\Gam$ (see Fig.\ref{f8tor}). Actually, the shape of this
spectrum is no different from the fluorescent spectrum emitted by
a non-polar  two-level  atom driven by a bichromatic field with
equal amplitudes of its components and symmetric detuning $\delta$
of their carrier frequencies from the atomic transition frequency
$\om_0$. Such a spectrum was predicted theoretically in
\cite{BichrTh1} and observed experimentally for the first time in
\cite{BichrExp}. A comprehensive theory of this phenomenon was
further developed\cite{BichrTh2,BichrTh3,BichrTh4}, where it was
shown that the spectrum of resonance fluorescence under symmetric
bichromatic excitation is made of well-separated peaks with the
central component at $\om_0$ and constant spacing $\delta$, where
$2\delta$ is the frequency difference between the two components
of the bichromatic driving field. The peak separations are
independent of the Rabi frequencies of these components, but the
number of observed peaks in the spectrum increases with the
increase of the Rabi frequencies, while their amplitudes
oscillate. Resonance-fluorescence and absorption spectra of a
two-level non-polar atom driven by a strong bichromatic field was
studied in \cite{F1}, while the case of the asymmetric excitation
of a two-level non-polar atom by a bichromatic field with one
strong and one weak component was addressed in \cite{F2}. For a
non-polar bichromarically driven two-level system  no visible
low-frequency fluorescence can be found. But in the case of the
polar system, when $\delta_a^{(1)}=\delta_a^{(2)} =\delta_a \ne
0$, this low-frequency spectrum consists of a series of peaks
separated by the same constant spacing $\delta$ as the
high-frequency spectral peaks (see Fig.\ref{f9tor}), the highest
of the peaks being centered at the Rabi frequency $\Om_R$. In this
particular respect and in its general appearance this spectrum is
strikingly similar to the central peak taken together with the
right sideband of the high-frequency spectrum in Fig.\ref{f8tor},
which, in its turn, remains virtually unchanged in the polar case
even for non-zero values of $\delta_a^{(1)}$ and $\delta_a^{(2)}$.
The amplitudes of the low-frequency peaks increase with increasing
$\delta_a$, while their spacings are independent of either
$\delta_a$ or $\Om_R$. The high-frequency spectrum also persists
without visible alterations for the non-polar or polar cases if
the system is driven by a trichromatic field with the third
low-frequency component being in resonance with the Rabi frequency
$\Om_R$ of the high-frequency bichromatic field but, at the same
time, $\delta_a^{(3)}=0$. The situation becomes totally different
for $\delta_a^{(3)}\ne 0$, wherein the high-frequency spectrum
retains its shape in general as to the number of the peaks, their
widths and positions (save some changes in the ratios of the peaks
amplitudes), whereas their amplitudes decrease in consistency with
increasing $\delta_a^{(3)}$. This complete evolution of the
fluorescent spectrum depending on the asymmetry parameter
$\delta_a^{(3)}$ can be seen in Fig.\ref{f10tor}. In some sense,
this effect is reminiscent of an optical transistor effect, in
that by changing the low-frequency radiation input one can alter
the character of the high-frequency fluorescence radiation output.
At the same time, the high-frequency fluorescence spectrum can be
altered to the very same effect not by increasing or decreasing
the low-frequency radiation input but rather by altering the
degree of the polar system asymmetry reflected in the magnitude of
the parameter  $\delta_a^{(3)}$. For example, for polar quantum
dot this can be done by altering its confining potential well
independently of the low-frequency radiation input. Therefore, it
seems justified to consider what is going on here not as an
optical, but rather as an optoelectronic transistor effect. Due to
the scalar product nature of the parameter $\delta_a^{(3)}$, one
more option to increase or decrease its magnitude is to alter the
direction of the low-frequency exciting field polarization in a
proper way. It is worth noticing in conclusion, that all the
spectra in the Fig.\ref{f10tor} remain virtually independent of
$\Om_R^{(3)}$, $\delta_a^{(1)}$ and $\delta_a^{(2)}$, only the
magnitude of the asymmetry parameter $\delta_a^{(3)}$  really
matters.

\section{Summary}

In this paper, the spectrum of  the  fluorescent radiation emitted
by a polar two-level system excited by a polychromatic field was
studied.  To summarize, we can conclude that by influencing a
two-level quantum polar system resonantly excited by a
high-frequency monochromatic laser field at the frequency of
atomic transition also by an auxiliary low-frequency monochromatic
field at the Rabi frequency of the high-frequency field, it is
possible to significantly transform the spectrum of the
high-frequency fluorescence of this quantum system. This effect
allows efficiently controlling the spectral properties of the
fluorescent radiation of the quantum system at high frequencies by
changing the intensity of the low-frequency component of the
bichromatic field thus formed. The control low-frequency field can
be significantly weaker than the high-frequency resonant field. It
was also shown that if this system is driven by a bichromatic
external field, with two high-frequency components of equal Rabi
frequencies and carrier frequencies being placed symmetrically
around the atomic transition frequency, and affected
simultaneously  by one more low-frequency component being in
resonance with the Rabi frequency of the high-frequency
components, then the spectrum of its fluorescent radiation can be
efficiently altered in a sensible way by controlling  the
intensity of the low-frequency field.  From this perspective, it
appears that the optimal candidate for the role of a polar quantum
system for the experimental detection and study of the effect of
the low-frequency component of a bichromatic field on the
high-frequency fluorescence spectrum is an isolated semiconductor
active quantum dot whose properties can be controlled by changing
the external electrostatic potentials applied to it, which expands
the experimental possibilities compared to the use of naturally
occurring polar quantum systems such as polar molecules. These
so-called artificial two-level atoms can possess much larger
matrix elements of the electric dipole moment with easily
controlled magnitudes than their naturally occurring atomic and
molecular counterparts\cite{Schulte:2015}. In particular, for
quantum dots, the values of the Rabi frequencies $\Om_R^{(j)}$ and
the parameters of violation of the inversion spatial symmetry
$\delta_a^{(j)}$, $j=1,...,N$ can be varied independently.
 From this perspective, asymmetric polar quantum dots look very promising
  and conducing not only from the strictly theoretical and
   experimental point of
view but rather in regard to practical applications of the effects
discussed in the present study.  And, moreover, the magnitudes of
all the Rabi frequencies $\Om_R^{(j)}$ and asymmetry parameters
$\delta_a^{(j)}$ can be adjusted and controlled independently of
each other. In particular, the effect of resonance fluorescence
from a single isolated semiconductor quantum dot was demonstrated
in \cite{Muller:2007, Unsleber:2015}, and somewhat later, the
effect of resonance fluorescence was used to generate a compressed
state of fluorescent radiation by a semiconductor quantum dot
\cite{Schulte:2015}.
 Further experimental and theoretical research in this domain of
quantum optoelectronics could facilitate development of useful
semiconductor nanodevices suitable for integration into much more
complex optoelectronic circuits.

\end{document}